\documentclass[prl,aps,twocolumn]{revtex4}

\usepackage{epsf}

\begin{document}

\title{%
\hfill{\normalsize\vbox{%
\hbox{\rm December 2006}
\hbox{\rm } }}\\
\vspace{-0.5cm}
{\bf Model for light scalars in QCD}}
\author{\addtocounter{footnote}{1}
{\bf Amir H. Fariborz}$^{\rm(a),\fnsymbol{footnote}}$}
\author{\addtocounter{footnote}{1}
{\bf Renata Jora}$^{\rm(b),\fnsymbol{footnote}}$}
\author{\addtocounter{footnote}{1}
{\bf Joseph Schechter}$^{\rm(b),\fnsymbol{footnote}}$}
\affiliation{$^{\rm(a)}$ Department of Mathematics/Science,
SUNY Institute of Technology, Utica, NY 13504}
\affiliation{$^{\rm(b)}$ Department of Physics, Syracuse University
Syracuse, NY 13244-1130}

\begin{abstract}
We propose a systematic procedure to study a 
 generalized linear sigma model
which can give a physical picture of possible mixing between $q{\bar q}$
and $qq{\bar q}{\bar q}$ low lying spin zero states. In the limit of 
zero quark masses, we derive the model independent results for the 
properties of the Nambu Goldstone pseudoscalar particles. For getting 
information on 
the scalars it is necessary to make a specific choice of terms. We 
impose two
plausible physical criteria - the modeling of the axial anomaly and the
suppression of effective vertices representing too many fermion lines -  
for limiting the large number of terms which 
are 
allowed on general grounds. We calculate the 
tree-level spectrum based on the leading 
terms in our approach
and find that it prominently exhibits a very low mass 
isosinglet scalar 
state. Finally we
 point out that the low energy result for scattering of pions  
 continues to hold in the general version of the model.

\end{abstract}

\maketitle

    Evidence has been accumulating \cite{ropp} for a very light mass
scalar-isoscalar particle, $f_0(600)$ as well as a possible similar
scalar-isospinor particle, $K_0^*(800)$. As has been widely discussed,
these may be joined with the well established scalars $a_0(980)$
and $f_0(980)$ to make a putative light scalar nonet. The upside down mass 
ordering of such a nonet suggests a four quark rather than a two quark
structure; both
  $qq$-${\bar q}{\bar q}$ \cite{j} 
and $q{\bar q}$-$q{\bar q}$ \cite{iw}
 forms have been proposed. 
Either alternative would be of great importance for a
full understanding of QCD in its non-perturbative low energy regime. 
The relation to the usual $q{\bar q}$ scalar mesons is of clear
relevance in a such a picture. It has been suggested
\cite{2-4mixing} that a mixing 
between the two quark and four quark nonets may help to better
understand certain anomalies of the two quark nonet spectrum.  
The resulting picture is complicated and one may wonder, for
example, whether the ordinary pions (believed to be of $q{\bar q}$
type) are chiral partners of the lighter four quark scalars or heavier
two quark scalars. Historically, it has been of great value to study
such questions in the framework of simple linear sigma models. Such a 
generalized linear sigma model
 was proposed in \cite{bfmns} and studied 
further in \cite{nr}
and \cite{fjs}. These papers have suggested the plausibility of a 
situation in which the lightest, approximate Nambu-Goldstone boson 
pseudoscalars are followed in ascending mass by scalars
with relatively large four quark content.
However, the model of interest may have many more terms
than previously considered; for example if 
the
interaction terms are restricted to be renormalizable, there are 
\cite{fjs}
21 chiral invariant terms and 21 additional terms with the chiral
transformation property of the QCD mass terms. In the present note we
attempt to understand the essential structure more clearly, to 
differentiate between model dependent and model independent results
as well as to suggest physical ways to choose the most important terms.
As an aid we first simplify the analysis by setting the light quark masses 
to zero. It is accepted that this is a reasonable qualitative 
approximation since the largest parts of the masses of all particles 
made of light quarks, other than the lightest $0^-$ octet, are expected
to  arise from 
spontaneous breakdown of chiral symmetry. 

   The fields of our ``toy" model consist of a 3 $\times$ 
3 matrix chiral nonet
field $M$, which represents $q{\bar q}$ type states as 
well as a
3 $\times$ 3 matrix chiral nonet
field $M^\prime$, which represents four quark 
type states.
      They have the
decompositions into scalar and pseudoscalar pieces:
$M = S +i\phi$,
$M^\prime = S^\prime +i\phi^\prime$ and
 behave under ${\rm SU(3)_L\times SU(3)_R}$ 
transformations  
as
 $ M \rightarrow U_L M U_R^\dagger$ and 
$M^\prime \rightarrow U_L M^\prime U_R^\dagger$.
However, 
 the ${\rm U(1)_A}$ transformation which acts at the 
quark level as $q_{aL}
 \rightarrow e^{i\nu} q_{aL}$, $q_{aR} \rightarrow e^{-i\nu} q_{aR}$
distinguishes the
two fields \cite{bfmns} according to
\begin{equation}
M \rightarrow e^{2i\nu} M, \hspace{1cm}
M^\prime \rightarrow e^{-4i\nu} M^\prime.
\label{MU1A}
\end{equation}
Note that our treatment is based only on the symmetry structure
and hence applies when $M'$ is any linear combination of
$qq$-${\bar q}{\bar q}$ 
and $q{\bar q}$-$q{\bar q}$ type fields.  
We will be interested in the situation where non-zero vacuum values
 of $S$ and $S'$ may exist: 
$\left< S_a^b \right> = \alpha \delta_a^b$,
$ \left< S_a^{\prime b} \right> =
\beta \delta_a^b$,
corresponding to an assumed ${\rm SU(3)_V}$ invariant
vacuum.
The Lagrangian density which defines our model is
\begin{eqnarray}        
{\cal L} = &-& \frac{1}{2} {\rm Tr}
 \left( \partial_\mu M \partial_\mu M^\dagger
 \right) - \frac{1}{2} {\rm Tr}
 \left( \partial_\mu M^\prime \partial_\mu M^{\prime \dagger} \right)
\nonumber \\
 &-& V_0 \left( M, M^\prime \right) - V_{SB},
\label{mixingLsMLag}
\end{eqnarray}
where $V_0(M,M^\prime) $ stands for a general function made
 from ${\rm SU(3)_L \times SU(3)_R}$
(but not necessarily ${\rm U(1)_A}$) invariants                              
formed out of
$M$ and $M^\prime$. The quantity $V_{SB}$ which represents
the effective chiral symmetry
breaking light quark mass terms
will be set to zero here.

  The ${\rm U(1)_A}$ transformation, which plays a 
special
role in this model,
suggests another useful simplification. In QCD there is a
special instanton induced term- the ``'t Hooft
determinant" \cite{t}- which breaks the ${\rm U(1)_A}$
symmetry and can be modeled as
${\rm det} (M)+ {\rm det} (M^\dagger)$. It thus may be 
natural to require all the 
terms to satisfy ${\rm U(1)_A}$ invariance except for a
 particular subset which could model
the ${\rm U(1)_A}$ anomaly. 
 If one demands 
that, as in QCD, the  ${\rm U(1)_A}$ variation of the 
effective Lagrangian be proportional to the gluonic
 axial anomaly then a similar effective term
${\cal L}_\eta = - c_3 
[{\rm ln} ({\rm det} (M)) - 
{\rm ln} ({\rm det} (M^{\dagger}))]^2$,
where $c_3$ is a numerical parameter,   
 seems appropriate \cite{ln}.

    A lot of information concerning especially 
the pseudoscalar
particles in the model may be obtained in general
 without even specifying
the terms in the potential. This may be achieved by
studying ``generating" equations which arise from
the demand that the infinitesimal symmetry transformations
in the model hold. For the nine axial transformations
one finds \cite{fjs}:
\begin{eqnarray}
&&[\phi,\frac{\partial V_0}{\partial S}]_+ -
[S,\frac{\partial V_0}{\partial \phi}]_+ +
(\phi,S)\rightarrow(\phi',S')=
\nonumber \\
&&1 [ 2 {\rm Tr} (\phi'\frac{\partial V_0}{\partial S'}-
S'\frac{\partial V_0}{\partial \phi'}) 
- 8 c_3 i {\rm ln} 
( \frac{{\rm det} M}{{\rm det} M^{\dagger}})].
\label{geneq}
\end{eqnarray}
To get constraints on the particle masses we will differentiate
these equations once with respect to each of the two matrix fields
$\phi$ and $\phi'$ and evaluate the equations in the
 ground state, taking into account the ``minimum" conditions,
$\langle \frac{\partial V_0}{\partial S} \rangle =0$ and
$\langle \frac{\partial V_0}{\partial S'} \rangle =0$. 
Further differentiations
with respect to all four matrix fields will
similarly yield ``model independent"
information on 3 and 4 point vertices.
We also require the Noether currents,
$(J_\mu^{axial})_a^b =\alpha\partial_\mu\phi_a^b +
\beta\partial_\mu{\phi'}_a^b+ \cdots$,
where the dots stand for terms bilinear in the fields.   
Using Eq.(\ref{geneq}) the squared mass matrix which mixes the
degenerate two quark
and degenerate four quark pseudoscalar
octets is:
\begin{equation}
(M_\pi^2)= y_\pi
\left[ \begin{array}{c c}
                \beta^2/\alpha^2  & -\beta/\alpha
\nonumber \\
                -\beta/\alpha  & 1
                \end{array} \right],
\label{mpipiprime}                                                  
\end{equation} 
where $y_\pi=\langle
{
  {\partial^2 V_0} \over {\partial {\phi'}_1^2 \partial {\phi'}_2^1}
}
\rangle$. Clearly, ${\rm det} (M_\pi^2)=0$ and the 
zero mass pion octet is a mixture of two quark and 
four quark fields.       
The transformation between the diagonal fields
 $\pi^+$ and $\pi'^+$ and the original pion fields
is defined as:
\begin{equation}
\left[
\begin{array}{c}  \pi^+ \\
                 \pi'^+
\end{array}
\right]
=
R_\pi^{-1}
\left[
\begin{array}{c}
                      \phi_1^2 \\
                        {\phi'}_1^2
\end{array}
\right]=
\left[
\begin{array}{c c}
                {\rm cos}\, \theta_\pi & - {\rm sin}\, 
\theta_\pi
\nonumber               \\
 {\rm sin}\, \theta_\pi & {\rm cos}\, \theta_\pi
\end{array}
\right]
\left[
\begin{array}{c}
                        \phi_1^2 \\
                        {\phi'}_1^2
\end{array}
\right],
\label{mixingangle}
\end{equation}
which also defines the transformation matrix, $R_\pi$.
The explicit diagonalization yields:
\begin{equation}
{\rm tan}\, {\theta_\pi}=-\frac{\beta}{\alpha},
\label{thetapi}
\end{equation}
which may be
interpreted as the ratio of the four quark condensate
to the two quark condensate in the underlying QCD. We see that the 
mixing between the two quark pion and the four
quark pion would vanish if the 
four quark condensate were to vanish in this model.
Rewriting the Noether current as $(J_\mu^{axial})_1^2 =F_\pi\partial_\mu 
\pi^+ + F_{\pi'}\partial_\mu
\pi'^+
+\cdots$ shows that                 
\begin{equation}
F_\pi=2\sqrt{\alpha^2+\beta^2},
\hspace{1cm}
F_{\pi^\prime}=0.
\label{nomassFpis}
\end{equation}                                                         
Note that the physical higher mass pion state decouples from the axial 
current.   Altogether the (eight) zero mass pseudoscalars
 are characterized by the three parameters $\alpha$, $\beta$ and
 $y_\pi$. On the other hand, there are only two experimental inputs:
$F_\pi$ = 131 MeV and the mass of $\pi(1300)$, the presumed higher mass 
pion candidate. Thus, 
the interesting question of the four quark content of the pion 
has an inevitably model dependent answer.

    Next let us consider the ``model independent" information available
for the two pseudoscalar SU(3) singlet states. This sector is related to 
the QCD axial anomaly. In the single M model, the anomaly can be modeled 
by the term ${\cal L}_\eta$ mentioned above. In the 
$M$-$M'$ model under 
consideration this form is no longer unique and it is natural to consider
a generalization \cite{genu1} in which 
${\rm ln} ( {\rm det} (M)/{\rm det} (M^\dagger) )$ is 
replaced by
$\gamma_1[{\rm ln} ({\rm det} (M)/{\rm det} 
(M^\dagger))]+(1-\gamma_1)[
{\rm ln} ({\rm Tr} (MM'^\dagger)/
{\rm Tr} (M'M^\dagger))]$, 
where $\gamma_1$ is a 
dimensionless parameter.
Then the squared mass matrix which mixes the two SU(3) pseudoscalar 
singlet states is obtained as:
\begin{equation}
(M^2_0)=
\left[ \begin{array}{c c}
                -\frac{8c_3(2\gamma_1+1)^2}{3\alpha^2} +z_0^2 y_0 & -z_0 
y_0+\frac{8c_3(1-\gamma_1)(2\gamma_1+1)}{3\alpha\beta}
\nonumber \\
                -z_0 y_0 
+\frac{8c_3(1-\gamma_1)(2\gamma_1+1)}{3\alpha\beta}  & 
y_0-\frac{8c_3(1-\gamma_1)^2}{3\beta^2}
                \end{array} \right] .
\label{phizeromixing}
\end{equation}
Here $z_0=-2\beta/\alpha$ and
$y_0=\langle \frac{\partial^2 
V}{\partial\phi'_0\partial{\phi'}_0}\rangle$.
Note that when $c_3$ is set to zero, making the entire
Lagrangian ${\rm U(1)_A}$ invariant, ${\rm det} (M_0^2)=0$. Then 
one
of the singlet pseudoscalars becomes, as well known, a
Nambu Goldstone particle. This occurs in the large number
of colors limit but 
we will not make that approximation here.

    In order to get information about the scalar
meson masses and mixings as well as to complete the
description of the 
pseudoscalars it is necessary to make a specific
choice of interaction terms. To proceed in a systematic way
we define the following quantity for each term,
\begin{equation}
N=2n+4n',
\label{count}
\end{equation}
where $n$  and $n'$ are respectively the number of $M$ fields 
and the number of $M'$ fields contained in that term. We shall
restrict our choice to the lowest non-trivial value of $N$,
which corresponds physically to the total number of quark
and antiquark lines at each vertex. In addition to the two
special terms which saturate the ${\rm U(1)_A}$ anomaly 
already mentioned,
this gives the leading ($N$=8)
potential
\begin{eqnarray}
V_0 =&-&c_2 \, {\rm Tr} (MM^{\dagger}) +
 c_4^a \, {\rm Tr} (MM^{\dagger}MM^{\dagger})
\nonumber \\
&+& d_2 \,
{\rm Tr} (M^{\prime}M^{\prime\dagger})
     + e_3^a(\epsilon_{abc}\epsilon^{def}M^a_dM^b_eM'^c_f + h.c.)
\nonumber \\
&+&\cdots,
\label{SpecLag}
\end{eqnarray}
where the dots stand for the ${\rm U(1)_A}$ non-invariant 
terms.
For simplicity, we have neglected the $N$=8 term,
 $c_4^b[{\rm Tr} (MM^\dagger)]^2$ which is suppressed, in the single $M$
model by the quark line rule.                                                         
It may be noted that the quantities ${\rm det} (M)$ and 
${\rm Tr} (MM'^\dagger)$
which enter into those two terms which saturate the 
${\rm U(1)_A}$ anomaly have $N$=6. In this counting 
scheme, 
${\rm U(1)_A}$ invariant terms with $N$=12 (and higher) 
might be successively 
added to improve the approximation. The minimum equations for this
potential are:
\begin{equation}
\left\langle { {\partial V_0} \over {\partial S_a^a} } \right\rangle =
2 \,\alpha\,  \left[ - c_2 + 2\, c_4^a\, \alpha^2 + 4\, e_3^a \, \beta 
\right] 
=0,
\label{mealpha}
\end{equation}
\begin{equation}
\left\langle { {\partial V_0} \over
{\partial {S'}_a^a} } \right\rangle
=
2  \left[ d_2\, \beta + 2\, e_3^a\, \alpha^2 \right] = 0.
\label{mebeta}
\end{equation}
The ${\rm U(1)_A}$ violating $c_3$ terms do not 
contribute to 
these equations.                                           
  Notice that $\alpha$ is an overall factor in Eq. (\ref{mealpha})
so that, in addition to the physical spontaneous breakdown solution where
$\alpha \ne 0$ there is a solution with $\alpha= 0$. On the other
hand, $\beta$ is not an overall factor of Eq. (\ref{mebeta}) and it is
easy to see that $\beta$, which measures the ``4 quark condensate", is 
necessarily non-zero in the physical
situation where $\alpha$ is non-zero.
From the specific form in Eq.(\ref{SpecLag}) we find the
mixing squared mass matrices for the degenerate octet scalars
as well as the SU(3) singlet scalars:
\begin{equation}
(X_a^2) =
\left[
\begin{array}{cc}
2 \, \left[- c_2 + 6\, c_4^a\, \alpha^2 - 2\, e_3^a \, \beta \right] &
-4\alpha e_3^a  \\
-4\alpha e_3^a  &
2d_2
\end{array}
\right]
\label{xa}
\end{equation}                             

\begin{equation}
(X_0^2) =
\left[
\begin{array}{cc}
2 \, \left[- c_2 + 6\, c_4^a\, \alpha^2 + 4\, e_3^a \, \beta \right] &
8\alpha e_3^a  \\
8\alpha e_3^a  &
2d_2
\end{array}
\right]
\label{x0}
\end{equation}                 
    Now let us consider the comparison of the model with experiment.
To start with there are 8 parameters ($\alpha$, $\beta$, $c_2$, $d_2$,
$c_4^a$, $e_3^a$, $c_3$ and $\gamma_1$). The last two parameters appear 
only in the mass matrices of the pseudoscalar  SU(3) singlets and
are conveniently discussed separately. The other six are effectively 
reduced to four 
by using the two minimum equations (\ref{mealpha}) and (\ref{mebeta}).
As the corresponding four experimental inputs
\cite{ropp} we take the non-strange
quantities:
\begin{eqnarray}
m(0^+ {\rm octet}) &=& m[a_0(980)] = 984.7 \pm 1.2\, {\rm MeV}  
\nonumber \\  
m(0^+ {\rm octet}') &=& m[a_0(1450)] = 1474 \pm 19\, {\rm MeV}  
\nonumber \\
m(0^- {\rm octet}') &=& m[\pi(1300)] = 1300 \pm 100\, {\rm MeV}  
\nonumber \\
F_\pi &=& 131 \, {\rm MeV}
\label{inputs1}
\end{eqnarray}
Evidently, the largest experimental uncertainty
 appears in the mass of $\pi(1300)$; we shall consider the other
masses as fixed at their central values and vary this mass in the 
indicated range. From
studying the scalar SU(3) singlet states we find the 
consistency
condition for positivity of the eigenvalues
 of their squared mass matrix, 
Eq.(\ref{x0}): 
\begin{equation}
m[\pi(1300)]< 1302\,{\rm MeV}.
\label{scalarconscond}
\end{equation}
The model predicts, as $m[\pi(1300)]$ varies from 1200 to 1300 MeV,
\begin{eqnarray}
m(0^+ {\rm singlet})&=& 510 \to 
28\hspace{.1cm} {\rm MeV},
\nonumber \\
m(0^+{\rm singlet}') &=&1506 
\to 1555\hspace{.1cm} {\rm MeV}.
\label{scalsing}
\end{eqnarray}
Clearly, the most dramatic feature is the very low mass of
the lighter SU(3) singlet scalar meson. Of course, one expects the 
addition
of quark mass type terms to modify the details somewhat.

   To calculate the masses of the SU(3) singlet pseudoscalars
we must diagonalize Eq.(\ref{phizeromixing})
 with the specific choices of parameters $y_0=2d_2$ and
$z_0=4e_3^a\alpha/d_2$ corresponding to the potential of
 Eq.(\ref{SpecLag}). This enables us to fit in principle,
for any choice of $m[\pi(1300)]$, the two
parameters $c_3$ and $\gamma_1$ in terms 
of the experimental masses of  $\eta$(958) and one of
the candidates $\eta(1295)$, $\eta(1405)$, $\eta(1475)$
and $\eta(1760)$. However, it turns out that the 
positivity of the eigenvalues of
the matrix $(M_0^2)$ imposes additional constraints on the choice of
 $m[\pi(1300)]$ in Eq.(\ref{scalarconscond}).
 Furthermore the first two candidates for 
the heavier 
$\eta$ are also ruled out on grounds of this positivity.
For $\eta(1475)$ the allowed range of $m[\pi(1300)]$ is restricted
to 1200 to 1230 MeV. On the other hand, there is no additional restriction 
if $\eta(1760)$ is chosen. If the choice of $\eta(1475)$ is made,
the predicted range of $m(0^+ {\rm singlet})$ is narrowed 
from that given
in Eq.(\ref{scalsing}) to $510\to 410$ MeV. 

\begin{figure}[htbp]
\begin{center}
\epsfxsize = 7.5cm
\ \epsfbox{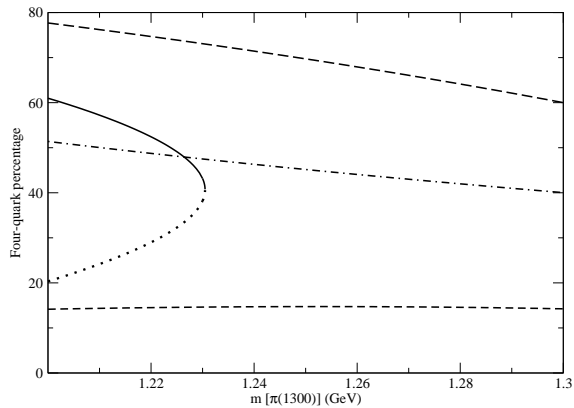}
\end{center}
\caption[]{%
Four quark percentages of the pion (dashed line), the $a_0(980)$
 (top long-dashed line), the very light $0^+ {\rm singlet}$ (dotted-dashed 
line)  and the
$\eta(958)$ in the scenario where the higher state is identified as the 
$\eta(1475)$ (curve containing both solid and dotted pieces)
as functions of the undetermined input parameter, 
$m[\pi(1300)]$. Note that there are two solutions for the $\eta(958)$: the 
dotted curve choice
gives it a predominant two quark structure and the solid curve choice,
a larger four quark content.  
}
\label{fig:1}
\end{figure}

    It is very interesting to see what the model has to say about the 
four 
quark percentages of the particles it describes. These percentages are 
displayed in Fig.\ref{fig:1} as functions of the precise value of the 
input parameter $m[\pi(1300)]$. The pion four quark content (equal
to 100 ${\rm sin} ^2\theta_\pi$) is seen to be about 17 
percent. Of course the 
heavier pion would have about an 83 percent four quark content. On the 
other hand, the octet scalar states present a reversed picture: the
$a_0(980)$ has a large four quark content while the $a_0(1450)$ has a 
smaller four quark content. The very light and the rather heavy
$0^+$ singlets are about maximally mixed, having roughly equal 
contributions from the 4 quark and 2 quark components. 

   The perhaps more 
plausible scenario in the case of the $0^-$ singlets takes  
 $\eta(1475)$ as the heavy 
$0^-$ singlet state. Fig. \ref{fig:1} shows that there are two solutions
for each value of $m[\pi(1300)]$; the dotted line gives a mainly 
$q{\bar q}$ content while the solid line gives a mainly four quark 
content. 
 Note that this scenario 
does not allow $m[\pi(1300)]$ to be higher than about 1230 MeV.   
The choice (not shown) of $\eta(1760)$ as the partner of $\eta(958)$
also leads to two solutions with small and large two quark content. 

    There are two reasons for next briefly discussing the pi-pi
scattering in this model. First, since the iso-singlet scalar
resonances above are being considered at tree level, one expects,
as can be seen in the single $M$ model also discussed in \cite{bfmns}
and at the two flavor level in \cite{as},
that unitarity corrections for the scattering amplitudes will
 alter their masses and widths. Second, since the pion looks
unconventional in this model (having a non-negligible four quark
component) one might worry that the fairly precise ``current algebra" 
formula for the near to threshold scattering amplitude might acquire
unacceptably large corrections. 
In the present massless pion model,
this formula \cite{w} should read,
\begin{equation}
A(s,t,u)=2s/F_{\pi}^2,
\label{caf}
\end{equation}
where $A(s,t,u)$ is the conventional amplitude term expressed in terms of
the Mandelstam variables $s$, $t$ and $u$. To obtain $A(s,t,u)$ 
in the present model, one needs 
the four point vertices involving the pseudoscalar octet fields as well
as the three point vertices involving two pseudoscalar octet fields and
one scalar field. It turns out \cite{fjssoon}
 that the result Eq.(\ref{caf}) follows
in a ``model independent" way just by using the generating 
Eq.(\ref{geneq}): the four point vertices can be related to the three
point vertices, which can in turn be related to the two point
vertices (masses). For example, the three point vertices involving the 
SU(3) singlet scalars can be related to the scalar and pseudoscalar
squared mass matrices as:
\begin{eqnarray}
&\frac{\sqrt{3}F_\pi}{2}\sum_{B}(R_\pi^{-1})_{1B}
\langle{{\partial^3V_0}\over{\partial({\phi}_1^2)_A\partial({\phi}_2^1)_B
\partial(S_0)_H}}\rangle
\nonumber \\
& = (X_0^2)_{AH}-(M_\pi^2)_{AH}.          
\label{3to2}
\end{eqnarray}
Here the capital Latin subscripts refer to summation over the unprimed
and primed fields, $(M_\pi^2)$ is given in Eq.(\ref{mpipiprime}) and 
$(X_0^2)$
is the model independent version of Eq.(\ref{x0}). It is interesting to 
note 
that the current algebra theorem  will ``tolerate" any amount of
four quark component in the massless pion. The present model clearly shows
that while the (lighter) pion is mainly two quark, the lighter scalars 
have very 
large four quark components. This is perhaps the opposite of what one 
might initially think and is related to the characteristic mixing pattern
emerging in a transparent form here.

   It is a pleasure to thank A. Abdel-Rehim, D. Black, M. Harada, S. 
Moussa, S. Nasri and F. Sannino for helpful discussions.
The work of A.H.F. has been partially supported by  
2005-2006 Faculty Grant Program, Office of VPAA, 
SUNY Institute of Technology. 
The work of R.J. and J.S. is supported in part by the U. S. DOE under
Contract no. DE-FG-02-85ER 40231.

\end{document}